\documentclass[iop,tighten]{emulateapj}
\usepackage{amsmath,epsfig,color}
\usepackage[breaklinks,colorlinks,citecolor=blue]{hyperref}
\begin{document}
\title{Implications from GW170817 for $\Delta$-isobar admixed 
hypernuclear compact stars}
\author{Jia Jie Li}
\email{jiajieli@itp.uni-frankfurt.de}
\affiliation{Institute for Theoretical Physics,
J. W. Goethe University,
D-60438 Frankfurt am Main, Germany}
\author{Armen Sedrakian}
\email{sedrakian@fias.uni-frankfurt.de}
\affiliation{Frankfurt Institute for Advanced Studies,
D-60438 Frankfurt am Main, Germany}
\affiliation{Institute of Theoretical Physics, University of Wroclaw,
50-204 Wroclaw, Poland}

\begin{abstract}
  The effects of $\Delta$ isobars on the equation of state  of 
  dense matter and structure of compact stars (CSs) are explored within
  the covariant density functional theory and confronted with 
  the data on tidal deformability (TD) extracted from the GW170817
  event. We show that the presence of $\Delta$ isobars substantially
  softens the tension between the predictions of the hypernuclear 
  density functionals and the inference from the observations of 
  relatively small radius and small TD of canonical mass CSs. 
  The TDs deduced from GW170817 are compatible with 
  the existence of hypernuclear CSs containing an admixture of 
  $\Delta$ isobars. We thus argue that the GW170817 event is 
  consistent with a merger of a binary CS system having both 
  strangeness (hyperons) and $\Delta$ isobars in the stellar core.
\end{abstract}

%---------------------------------------------------------------
\section{Introduction}
\label{sec:intro}
%---------------------------------------------------------------

The first multi-messenger observations of gravitational waves (GW)
from the binary neutron star (NS) merger event, GW170817, marks the
start of a new era in astronomy and
astrophysics~\citep{LIGO_Virgo2017b,
  LIGO_Virgo2017c,LIGO_Virgo2017a}. The significance of this and
future similar events lies, in part, in the opportunity of gaining
insight into the equation of state (EoS) and composition of dense
matter~\citep{LIGO_Virgo2017c,LIGO_Virgo2018a,LIGO_Virgo2018b,
  Margalit2017,Bauswein2017,Ruiz2018,Rezzolla2018,Soares-Santos2017,Villar2017}.
In particular, the TD (or polarizability) of NSs deduced from GW170817
puts already additional constraints on stellar radii and ultimately on
the details of the EoS~\citep{Fattoyev2018,Annala2018,
  Desoumi2018,Most2018,Paschalidis2018,Tews2018,Zhangnb2018}.

The potential implications of the GW170817 event cover a wide range
of fundamental questions associated with NSs, ranging from their 
interior composition to their role in nucleosynthesis. In fact,
a number of works have explored the possibility of using the
observations of GW170817 to probe the occurrence of a hadron-quark
phase
transition~\citep{Paschalidis2018,Most2018,Burgio2018,Blaschke2018,Alvarez2019}.

Even at the level of the hadronic matter the composition of stellar
matter could be rather complex. There have been intense studies of a
number of possibilities of new degrees of freedom, such as
hyperons~\citep{Glendenning1985,Prakash1992,Weissenborn2012,
  Dalen2014,Oertel2015,Tolos2016,Fortin2017,Lijj2018a}, $\Delta$
isobars~\citep{Prakash1992,Schurhoff2010,Drago2014,Caibj2015,
  Kolomeitsev2017,Lijj2018b} or other exotic hadronic states such as
$d^*(2380)$ resonance~\citep{Vidana2018}. The existence of hyperons
inside NSs has been questioned for a long time because a class of
models based on non-relativistic microscopic treatments of dense
matter were not able to produce massive enough ($M\sim 2M_{\odot}$)
stars.  However, the hyperonization cannot be simply ruled out by the
existence of $2M_\odot$ stars. In particular, covariant density functional (CDF) based models are
versatile enough to resolve the problem by tuning the interactions in
the hyperonic sector to hypernuclear data and CS
masses~\citep{Dalen2014,Oertel2015,Tolos2016,Weissenborn2012,
  Tolos2016,Fortin2017,Lijj2018a}. Of course, one may also conjecture
a layer of hypernuclear matter in-between the nucleonic and quark
matter phases~\citep{Bonanno2012,Masuda2013,Zdunik2013,Dexheimer2015}.
These models based on purely hyperonic EoS predict CS sequences with
$M_{\text{max}} \gtrsim 2.0M_\odot$ and radii of the canonical-mass
$M \sim 1.4M_\odot$ star $R_{1.4} \gtrsim 13$
km~\citep{Katayama2015,Fortin2016,Tolos2016,Lijj2018a,Lijiajie2019}.

The possibility of hypernuclear stars with small radii
($R_{1.4} \lesssim 13$~km) is as exciting as it is challenging for
nuclear theory. This requires a sufficiently soft EoS below 2-3
$\rho_{\text{sat}}$, where $\rho_{\text{sat}}$ is the nuclear
saturation density, while the observed large masses require that 
the same EoS must evolve into a stiff one at high densities. In 
this work, we construct such a model which is based on purely 
hadronic forms of stellar matter. We use the CDF theory~\citep{Mengjie2006}
to explore the effects of the $\Delta$ isobars on the TDs of CSs. 
We compare our results with the recent limits placed by the 
GW170817 event and discuss their implications for the interpretation
and detection of the current and future GW signals from NS merger.

%---------------------------------------------------------------
\section{Tidal deformability }
%---------------------------------------------------------------

Consider a static, spherically symmetric star, placed in a static
external quadrupolar tidal field of the companion. As two stars
approach each other during the early stages of an inspiral due to
their mutual gravitational attraction they experience tidal
deformation effects that can be quantified in terms of the TD
$\lambda$. It can be expressed in terms of the dimensionless tidal
Love number $k_2$ and the star's radius $R$ as
%---------------------------------------------------
$
\lambda = ({2}/{3})k_2R^5.
$
%---------------------------------------------------
The tidal Love number $k_2$ is calculated along with the solution
of the Tolman-Oppenheimer-Volkov equations~\citep{Hinderer2008,
Flanagan2008} and measures how easily the bulk of the matter in 
a CS is deformed~\citep{Hinderer2008,Flanagan2008,Binnington2009}. 
It is more convenient to work with the dimensionless TD $\Lambda$, 
which is related to the Love number $k_2$ and the compactness 
parameter $C = M/R$ through
%---------------------------------------------------
\begin{equation}
\Lambda = \lambda/M^5 = \frac{2}{3}\frac{k_2}{C^5}.
\end{equation}
%---------------------------------------------------
The total tidal effect of two CSs in an inspiraling binary
system is given by the mass-weighted TD
%---------------------------------------------------
\begin{equation}\label{eq:massweighted_lambda}
\tilde{\Lambda} = \frac{16}{13}\Bigg[\frac{(M_1 + 12M_2)M^4_1\Lambda_1}{(M_1 + M_2)^5}
+ 1 \leftrightarrow 2\Bigg],
\end{equation}
%---------------------------------------------------
where $\Lambda_1(M_1)$ and $\Lambda_2 (M_2)$ are the TDs of the
individual binary components. The quantity $\tilde{\Lambda}$ is
usually evaluated as a function of the chirp mass
$\mathcal{M} = (M_1 M_2)^{3/5}/M^{1/5}_T$ for various values of the
mass ratio $q = M_2/M_1$, where $M_T = M_1+M_2$ is the total mass of
the binary.

%---------------------------------------------------------------
\section{Hadronic matter equation of state}
%---------------------------------------------------------------
%
%----------------------------------------------------------
\begin{table*}[tb]
\caption{
  Properties of CSs assuming purely hyperonic ($Y$) and 
  hyperon-$\Delta$ admixed ($\Delta$) composition for selected 
  EoS models. The first three columns identify the EoS by the 
  isoscalar skewness $Q_{\text{sat}}$ (MeV) and isovector slope
  $L_{\text{sym}}$ (MeV). The remaining columns display: maximum
  mass $M_{\text{max}}$ ($M_\odot$), radius $R_{1.4}$ (km), 
  TDs $\lambda_{1.4}$ ($10^{36}$ gr cm$^2$ s$^2$) and $\Lambda_{1.4}$
  of canonical-mass CS. The results for hyperon-$\Delta$ admixed
  matter are obtained by tuning the $\Delta$-potential 
  $V_{\Delta}/V_N$ from 1 to 5/3.
}
\setlength{\tabcolsep}{10.0pt}
\label{tab:Cstars}
\begin{tabular}{cccccccccccc}
\hline
DF&$Q_{\text{sat}}$&$L_{\text{sym}}$& &$M_{\text{max}}^{(Y)}$&$R^{(Y)}_{1.4}$&$\Lambda^{(Y)}_{1.4}$ & &
$M^{(\Delta)}_{\text{max}}$ & $R^{(\Delta)}_{1.4}$ & $\lambda^{(\Delta)}_{1.4}$ & $\Lambda^{(\Delta)}_{1.4}$\\
\hline
1 & 480 & 40 & & 2.02 & 13.06 & 636 & & $2.03-2.06$ & $12.91-12.08$ & $3.25-1.97$ & $575-348$\\
2 & 480 & 60 & & 2.00 & 13.35 & 754 & & $2.01-2.04$ & $13.24-12.38$ & $3.97-2.42$ & $701-426$\\
3 & 300 & 50 & & 1.97 & 13.10 & 655 & & $1.98-2.01$ & $12.94-12.07$ & $3.30-1.94$ & $584-342$\\
4 & 800 & 50 & & 2.06 & 13.31 & 772 & & $2.07-2.10$ & $13.22-12.39$ & $4.08-2.56$ & $722-454$\\
\hline
\end{tabular}
\end{table*}
%-----------------------------------------------------------

Below we shall concentrate on the EoS of hypernuclear matter 
obtained from the CDF theory in its version
which uses density-dependent meson-baryon couplings~\citep{Typel1999}.
The extension to the hyperonic sector is described elsewhere~\citep{Lijj2018a,
Lijj2018b}, see also~\citep{Weissenborn2012,Fortin2017}.
A useful parameterization of the nucleonic EoS is given by 
the formula
%-----------------------------------------------------------
\begin{eqnarray}
\label{eq:EoS_param}
E(\rho,\delta) &\simeq & E_{\text{sat}} + \frac{1}{2!}K_{\text{sat}}n^2
                + \frac{1}{3!}Q_{\text{sat}}n^3 \nonumber \\
            & & + E_{\text{sym}}\delta^2
                + L_{\text{sym}}\delta^2n
                + {\mathcal O}(n^4,n^2\delta^2),
\end{eqnarray}
%-----------------------------------------------------------
where $n=(\rho-\rho_{\text{sat}})/3\rho_{\text{sat}}$ and
$\delta = (\rho_{\text{n}}-\rho_\text{p})/\rho$. The coefficients
entering the parameterization \eqref{eq:EoS_param} are the nuclear
{\it characteristic parameters} $E_{\text{sat}}$ (binding energy),
$K_{\text{sat}}$ (compressibility), $ E_{\text{sym}}$ (symmetry
energy), $Q_\text{sat}$ (isoscalar skewness coefficient) and
$L_\text{sym}$ (isovector slope coefficient), all defined at 
$\rho_{\text{sat}}$. The low-order characteristics $E_{\text{sat}}$,
$K_{\text{sat}}$ and $ E_{\text{sym}}$ are either strongly 
constrained or have no noticeable impact on the gross properties 
of CSs~\citep{Lijiajie2019,Margueron2018}. Therefore, the remaining
{\it higher-order characteristic parameters} are those that can
be used to calibrate the parameters of the density functional to 
the desired properties of the system at hand~\citep{Lijiajie2019}. 
Their values are weakly constrained by the conventional fitting 
protocol used in constructing the CDF. Specifically, their ranges 
can be constrained by $\chi$EFT computations of neutron matter~\citep{Drischler2016}
and the requirement that the EoS reproduces the observed maximum 
mass of a CS $M_{\text{max}}\gtrsim 2M_\odot$~\citep{Antoniadis2013}.
The {\it low-order characteristic parameters} for our models are
the same as the values predicted by the DD-ME2 parametrization~\citep{Lijj2018b}.
The observations~\citep{Antoniadis2013} and recent inferences of
$M_{\text{max}}$~\citep{Margalit2017,Ruiz2018,Rezzolla2018}, 
enable one to limit the range of $Q_\text{sat}$. However, it should
be stressed that such range will essentially depend on the composition
of matter~\citep{Lijiajie2019}. For instance, we have checked that 
the constraint $1.97 \lesssim M_{\text{max}}/M_\odot \lesssim 2.17$
limits  $Q_{\text{sat}}$ to the range $\sim [-650, -400]$ for
purely nucleonic matter, and $\sim [300, 800]$ for hyperonic matter.

In Table~\ref{tab:Cstars} we present integral parameters of hyperonic CSs
(maximum mass $M_{\text{max}}$, the radius $R_{1.4}$ and dimensionless
TD $\Lambda_{1.4}$ of canonical-mass star) for pairs of values of
$L_{\text{sym}}$ and $Q_{\text{sat}}$. To obtain massive enough
hypernuclear CSs, $Q_\text{sat}$ values must be
large~\citep{Lijiajie2019}; as a result the radii for canonical stars
obtained with these values of $Q_\text{sat}$ are $R_{1.4} \gtrsim 13$
km~\citep{Katayama2015,Fortin2016,Tolos2016,Lijj2018a,Lijiajie2019}.
The corresponding TDs are $\Lambda_{1.4} > 600$. To obtain
smaller values of these parameters, which are favored by the
observational data, we next explore the effect of $\Delta$ isobars by
varying the $\Delta$-potential. The underlying EoS is parameterized in
terms of the potential $V_\Delta$ of $\Delta$-isobar in symmetric
nuclear matter at $\rho_{\text{sat}}$.

To set the stage for the discussion of TDs of CSs in the next section
we show in Fig.~\ref{fig:Mass_Rad} the mass-radius (MR) relations for
purely nucleonic, hyperonic and hyperon-$\Delta$ admixed stellar
matter for a number of parameter values, as indicated in the
figure. It is seen that the allowance of $\Delta$-isobars in matter
shifts the radii of configurations to smaller values without affecting
seriously the value of the maximum mass
($M_{\text{max}} \gtrsim 2.0M_\odot$), an observation already made
using alternative parametrizations of CDF
in~\citep{Drago2014,Lijj2018b}.  We now turn to the discussion of our
results for TDs.

%---------------------------------------------------------------
\section{Results and Discussions}
%---------------------------------------------------------------

\begin{figure}[tb]
\centering
\ifpdf
\includegraphics[width = 0.45\textwidth]{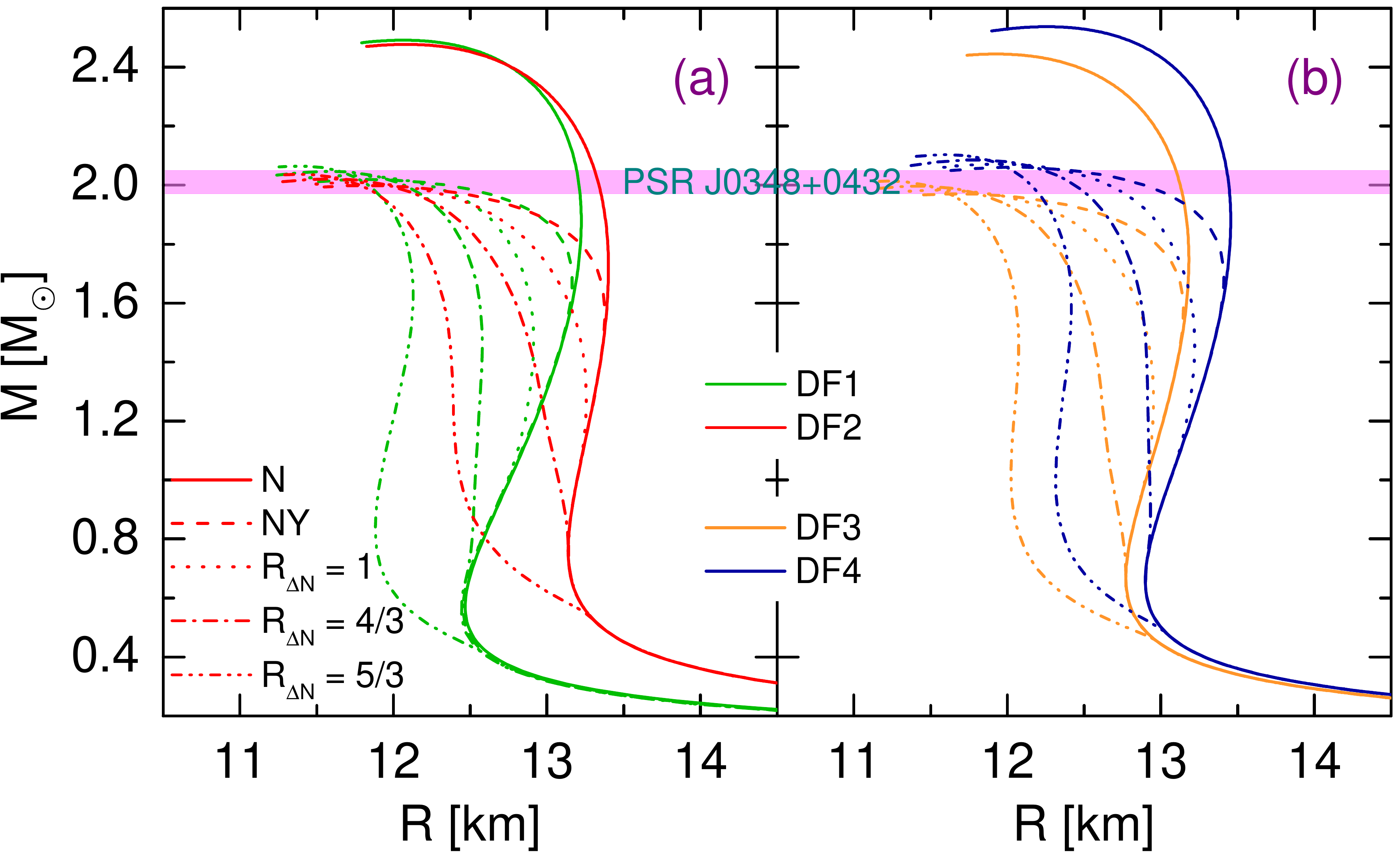}
\else
\includegraphics[width = 0.45\textwidth]{Mass_Rad.eps}
\fi
\caption{ Mass-radius relation for a set of EoS with varying
  $L_{\text{sym}}$ (a) and $Q_{\text{sat}}$ (b) and assuming 
  purely nucleonic ($N$), hyperonic ($NY$), and hyperon-$\Delta$
  admixed ($NY\Delta$) compositions of stellar matter. Three
  values of the $\Delta$-potential have been used:
  $R_{\Delta N} = V_\Delta/V_N =1$, 4/3 and 5/3, where $V_N$
  is the nucleon potential in isospin-symmetrical matter at 
  saturation density.}
\label{fig:Mass_Rad}
\end{figure}

The analysis of the GW170817 and complementary electromagnetic 
data could provide an important bound on the TDs. In particular, 
the LIGO/Virgo data analysis from GW170817 (hereafter LV
constraint) placed an upper bound for the case of low-spin 
priors~\citep{LIGO_Virgo2017c,LIGO_Virgo2018b}. We note that 
the LV detection of GW170817 derived an upper bound
$\tilde{\Lambda} \lesssim 900$~\citep{LIGO_Virgo2017c} from the
phase-shift analysis of the observed signal, and was recently
reanalyzed to be $\tilde{\Lambda} \lesssim 720$~\citep{LIGO_Virgo2018b}.
However, since this boundary is somewhat dependent on the waveform models,
we use here all the data reported in~\citep{LIGO_Virgo2017c,LIGO_Virgo2018b}.
A lower bound of $\tilde{\Lambda} \gtrsim 400$ is imposed by combining
optical/infrared and GW data with new numerical relativity
results~\citep{Radice2018}. More recently, an alternative approach
involving radiative transfer simulations for the electromagnetic
transient AT2017gfo predicts the lower bound to be
$\tilde{\Lambda} \gtrsim 197$~\citep{Coughlin2018}.
In the equal-mass scenario, these limits translate to constraints 
on the TD of a single CS itself.

\begin{figure}[b]
\centering
\ifpdf
\includegraphics[width = 0.45\textwidth]{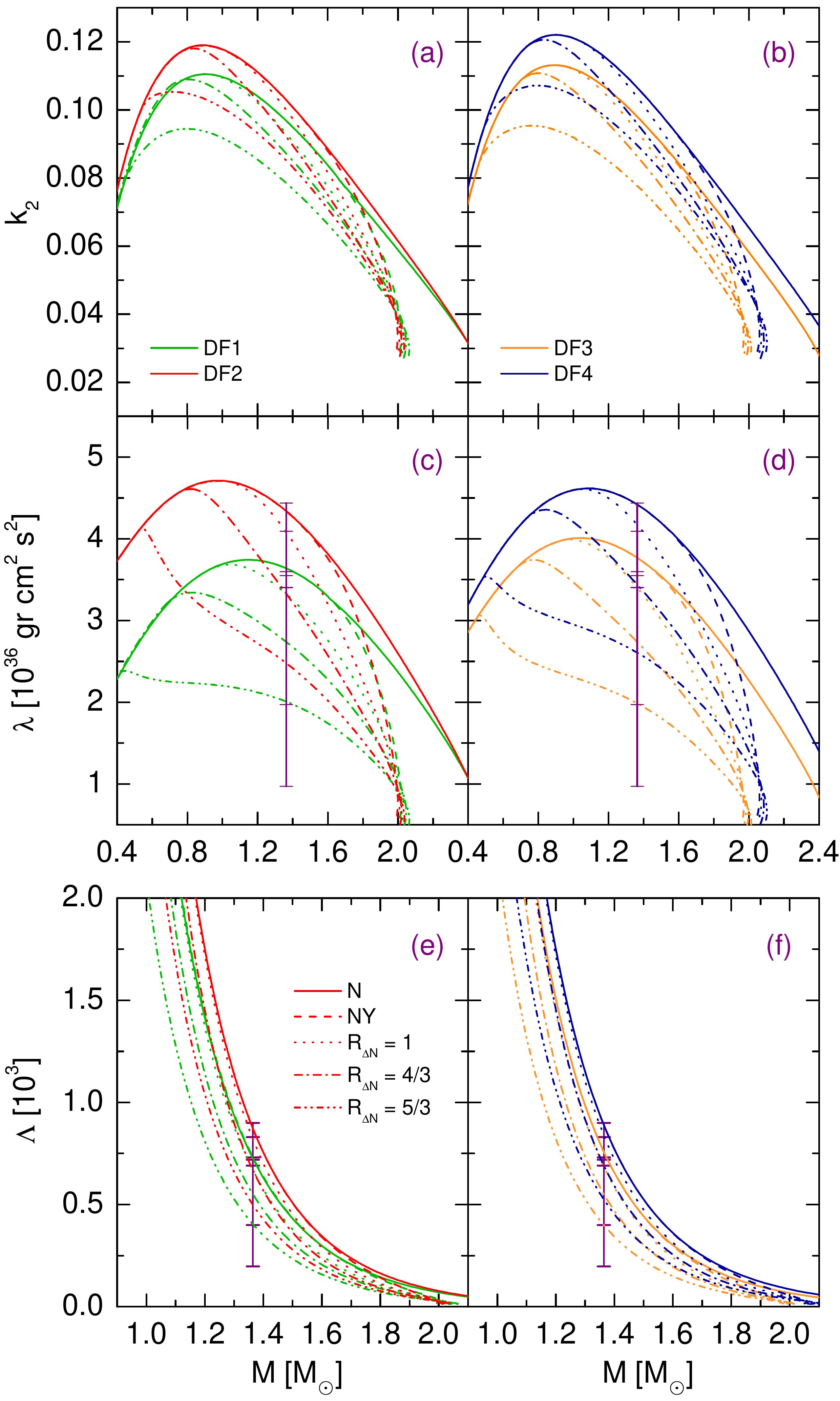}
\else
\includegraphics[width = 0.45\textwidth]{Lam_Mass.eps}
\fi
\caption{ Tidal Love number $k_2$ (top), polarizability $\lambda$
  (middle), and dimensionless TD $\Lambda$ (bottom)
  of a CS as a function of mass $M$ for a set of EoS with
  varying values of $L_{\text{sym}}$ (a, c, e), $Q_{\text{sat}}$ (b,
  d, f) and the $\Delta$-potential. The error bars indicate the
  constraints on $\Lambda (\lambda)$ for a $M = 1.362M_\odot$ star, 
  as estimated from the GW170817 event ~\citep{LIGO_Virgo2017c,
  LIGO_Virgo2018b,Radice2018,Coughlin2018}.
}
\label{fig:Lam_Mass}
\end{figure}

In Fig.~\ref{fig:Lam_Mass} we show the tidal Love number $k_2$, the
polarizability $\lambda$, and the dimensionless TD $\Lambda$ of a CS as 
a function of its mass $M$. It is seen that the value of $k_2$ peaks 
for stellar configurations with mass near $0.8M_\odot$, while it 
decreases rapidly for both higher and lower mass configurations. 
The behavior of $k_2$ as a function of $M$ as seen in Fig.~\ref{fig:Lam_Mass} (a, b)
can be understood by noting that the more centrally condensed 
stellar models have smaller $k_2$ values~\citep{Hinderer2010}.

The TD $\lambda$ for the same set of EoS is shown in 
Fig.~\ref{fig:Lam_Mass} (c, d). This parameter has a direct
astrophysical significance because it is proportional to a quantity
directly measurable by GW observations of inspiraling star binaries. 
As can be seen from Fig.~\ref{fig:Lam_Mass} (c, d), for each EoS
$\lambda$ follows a trend that is very similar to that of $k_2$. 
However, $\lambda$ is, in addition, proportional to $R^5$ and
thus it shows a more pronounced variation compared to $k_2$.
The error bars that are shown in Fig.~\ref{fig:Lam_Mass} (c, d) denote
the range of probable values of the TD $\lambda$ for a star 
with $M = 1.362M_\odot$. The relatively small $\lambda$
suggested by GW170817 implies that both the isoscalar and the
isovector sectors of the EoS up to 2-3$\rho_{\text{sat}}$ should
be soft. This, in turn, provides additional constraints for the 
nuclear {\it characteristic parameters}, in particular, the 
combinations of $Q_{\text{sat}}$ and $L_{\text{sym}}$. The 
dimensionless TD $\Lambda$ for selected EoS models is presented
in Fig.~\ref{fig:Lam_Mass} (e, f).

We now explore how the variations of $Q_{\text{sat}}$ and
$L_{\text{sym}}$ affect the TDs of CSs. We concentrate only on the
hyperonic models, but the conclusions also apply for $\Delta$-admixed
models. It is seen from Fig.~\ref{fig:Lam_Mass} (c, d) that the
variation of $L_{\text{sym}}$ from 40 to 60~MeV (with $Q_{\text{sat}}$
fixed) as allowed by the $\chi$EFT calculations has appreciable effect
on $\lambda$ for less massive stars ($M \lesssim 1.4M_\odot$), whereas the
variation of $Q_{\text{sat}}$ from 300 to 800~MeV (with
$L_{\text{sym}}$ fixed) as allowed by both the $\chi$EFT and the
maximum mass constraints has a more significant effect on $\lambda$ of
heavier stars ($M \gtrsim 1.4M_\odot$). The former observations on
$L_{\text{sym}}$ is consistent with previous
studies~\citep{Fattoyev2013}. Here we report, for the first time, a
broad analysis of the effects of variations of $Q_{\text{sat}}$.  The
stars of interest ($M \approx 1.1-1.6M_\odot$) are just at the
intersection where the effects of $Q_{\text{sat}}$ and
$L_{\text{sym}}$ on the TD are comparable. Therefore, $Q_{\text{sat}}$
or $L_{\text{sym}}$ values alone are insufficient to characterize the
low-density (up to $\sim 2\rho_{\text{sat}}$) behavior of EoS~\citep{Lijiajie2019}.
Further observations for binary merger events with smaller chirp mass
will allow one to narrow down the uncertainty in $L_{\text{sym}}$
(important for low-mass stars); larger chirp mass observations will
constrain more tightly $Q_{\text{sat}}$ (important for high-mass
stars).

Below we discuss results for a single set of EoS models with fixed
$Q_{\text{sat}}$ and $L_{\text{sym}}$, but the conclusions also apply
for other sets of models specified by these parameters. As already 
seen from the MR relation, accounting for $\Delta$ isobars reduces the 
radii of models from their values obtained for purely hyperonic 
(or nucleonic) EoSs. As expected, the deeper the potential $V_\Delta$, 
the larger is the observed shift in the radius is and, accordingly, 
the larger the reduction of the TD. For example, for $V_\Delta/V_N = 5/3$ 
the radius of a canonical $1.4M_\odot$ CS is by about 1~km smaller 
than the radius of its purely hyperonic (or nucleonic) counterpart.
Correspondingly, the dimensionless TD of a canonical $1.4M_\odot$CS
is by about 300 smaller than the value of its purely hyperonic
counterpart; see Fig.~\ref{fig:Lam_Mass} (e, f).

Table~\ref{tab:Cstars} shows how the properties of CS (including the
maximum mass, radius, and TD for a $1.4M_\odot$ star) change with the
value of the $\Delta$-potential $V_\Delta$.  The differences in the
values of $\Lambda_{1.4}$ can be mapped on the differences in the
compactness using the scaling $\Lambda \varpropto C^{-6}$ valid for
moderate-mass stars~\citep{Hinderer2010,Postnikov2010}. This scaling
follows from the proportionality $k_2 \varpropto C^{-1}$, which is
observed for a wide variety of EoS in the mass range
$1.1 \lesssim M/M_\odot \lesssim 1.6$ relevant for
GW170817. This mass range corresponds roughly to
$0.11 \lesssim C \lesssim 0.20$~\citep{Hinderer2010,Postnikov2010}.
We note that for the same mass stars (e.g. $M=1.4M_\odot$) the larger
is the $\Delta$-potential the smaller the radii are and, therefore,
the larger is the compactness and the smaller is the value of $\Lambda$.

We also observe in Fig.~\ref{fig:Lam_Mass} (c, d) that the mass
corresponding to the maximum value of $\lambda$ becomes smaller for
EoS models that have a deeper potential $V_\Delta$. As the increase
in the maximum masses of CSs due to $\Delta$ isobars is marginal,
it, more importantly, affects medium-mass stars. For large enough
$\Delta$-potentials the isobars may appear already at about
$2\rho_{\text{sat}}$, which implies that even low-mass CSs can be
affected by the populations of $\Delta$ isobars, see
Fig.~\ref{fig:Lam_Mass} (c, d). It is worth noticing that the
reduction of TD caused by the inclusion of $\Delta$ isobars in the
composition of matter is not very sensitive to the exact values of
$Q_{\text{sat}}$ and $L_{\text{sym}}$. However, the onset density of
$\Delta$ isobars and their fraction in the matter are sensitive to the
values of $Q_{\text{sat}}$ and $L_{\text{sym}}$.

\begin{figure}[tb]
\centering
\ifpdf
\includegraphics[width = 0.45\textwidth]{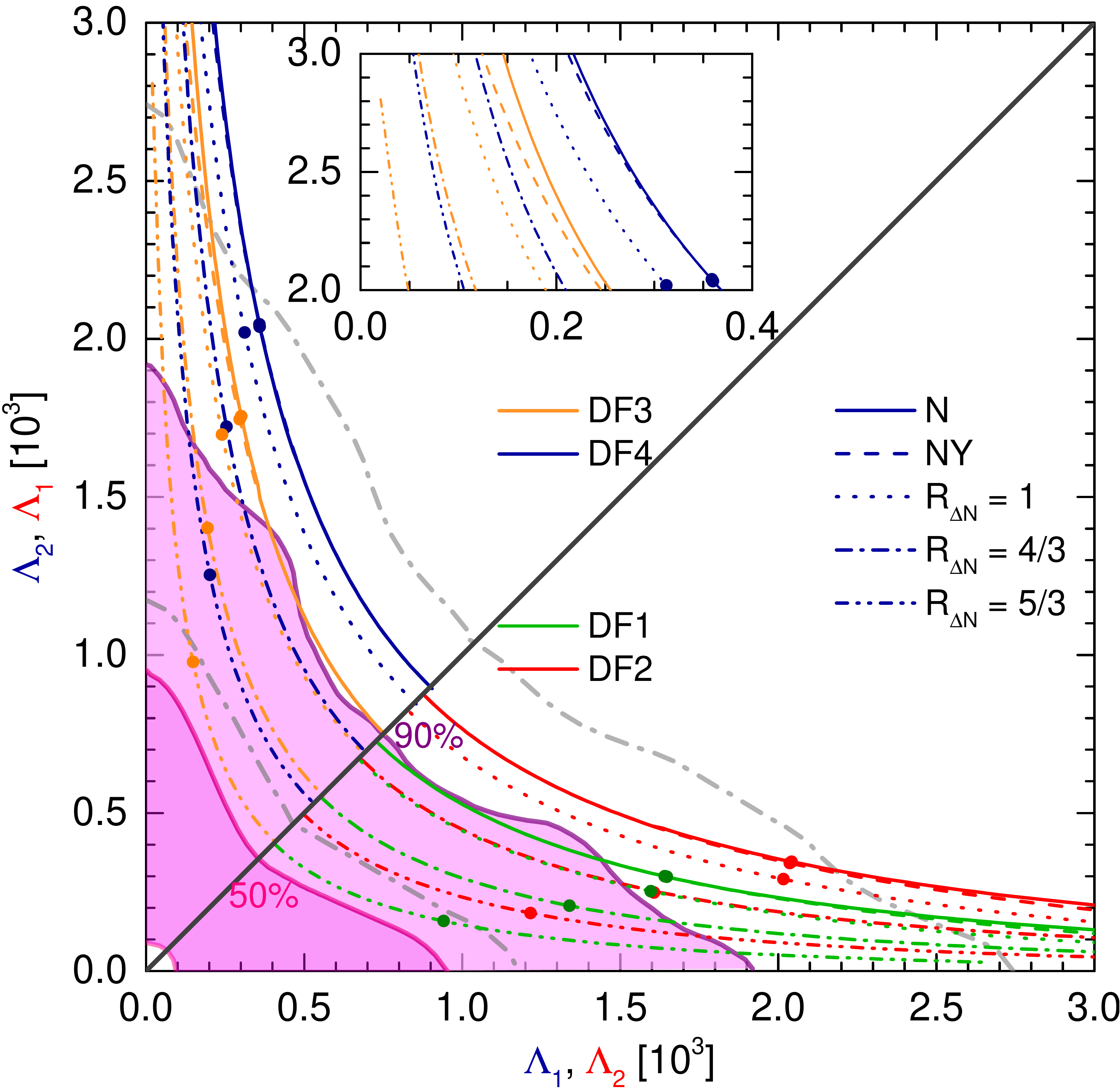}
\else
\includegraphics[width = 0.45\textwidth]{Lam_Lam.eps}
\fi
\caption{ Dimensionless TDs associated with the binary in GW170817,
  predicted by a range of EoSs that allows various the $L_{\text{sym}}$
  or $Q_{\text{sat}}$ individually, and the $\Delta$-potential. 
  The shadings correspond to the updated 50\% and 90\% credibility 
  regions (PhenomPNRT model) from the LV analysis~\citep{LIGO_Virgo2018b}. 
  The gray dashed-dotted curves represent the previously reported data 
  (TaylorF2 model)~\citep{LIGO_Virgo2017c}. The circles represent
  model predictions for a binary system having mass ratio $q = 0.73$.}
\label{fig:Lam_Lam}
\end{figure}

We now directly compare our modeling with the observational analysis
from the GW170817 event, assuming a chirp mass
$\mathcal{M} = 1.186 M_\odot$. We only investigate the more realistic
low-spin case, because large spins are not expected from the observed
galactic binary NS population~\citep{LIGO_Virgo2018a}.

In Fig.~\ref{fig:Lam_Lam} we display predictions from all EoS models
for the individual tidal polarizabilities $\Lambda_1$-$\Lambda_2$
associated with the $M_1$-$M_2$ components of the binary. The diagonal
line corresponds to the case of an equal-mass binary, i.e., 
$M_1 = M_2 = 1.362M_\odot$. The shaded areas correspond to the 90\% 
and 50\% confidence limits, which are obtained from the improved 
analysis of the GW170817 event~\citep{LIGO_Virgo2018b}. The previously
reported data~\citep{LIGO_Virgo2017c} are also shown as they have
been widely used to constrain EoS models.

It is clearly seen from Fig.~\ref{fig:Lam_Lam} that all the hyperonic
(or nucleonic) EoS models satisfy the original 90\%
confidence~\citep{LIGO_Virgo2017c}, whereas the two sets of hyperonic
EoS models, $(Q_{\text{sat}}, L_{\text{sym}}) = (800, 50)$ and
$(480, 60)$ [MeV], are ruled out by the updated
data~\citep{LIGO_Virgo2018b}. The remaining two sets of hyperonic EoS
models with $(Q_{\text{sat}}, L_{\text{sym}}) = (300, 50)$ and
$(480, 40)$ [MeV], closely follow the updated 90\%
confidence~\citep{LIGO_Virgo2018b}. As soon as the $\Delta$ isobars
appear in matter with reasonably attractive $\Delta$-potential in
nuclear matter ($V_\Delta < V_N$) all EoS models are completely
inside the region of compatibility with the data from GW170817. 
Thus, if the binary in the GW170817 event contained hadronic stars, 
the updated analysis would strongly favor the presence of $\Delta$ isobars
in addition to hypernuclear matter, as our purely hyperonic EoS 
models are rather representative of this class of models. An
alternative is the strong phase transition to a quark matter phase,
as it has been shown that such a transition also helps in producing 
more compact objects than those composed of nucleons 
only~\citep{Kojo2016,Alford2017,Blaschke2018,Alvarez2019,Montana2018,Xia2019}.

Interestingly, the $\Lambda_1$-$\Lambda_2$ curves predicted by EoS
models in the parameter spaces
$(Q_{\text{sat}}, L_{\text{sym}}, V_\Delta) = (480, 40, V_N)$ and
$(480, 60, 4/3V_N)$~MeV are almost identical, but their MR relations
are noticeably different. This is because the two MR curves cross
around $M = 1.4M_{\odot}$, and converge to each other at
$M \simeq 2.0M_{\odot}$, see Fig.~\ref{fig:Mass_Rad} (a). 
It is thus possible that different $M_1-M_2$ components
of the binary have almost the same polarizabilities $\Lambda_1-\Lambda_2$.

\begin{figure}[tb]
\centering
\ifpdf
\includegraphics[width = 0.45\textwidth]{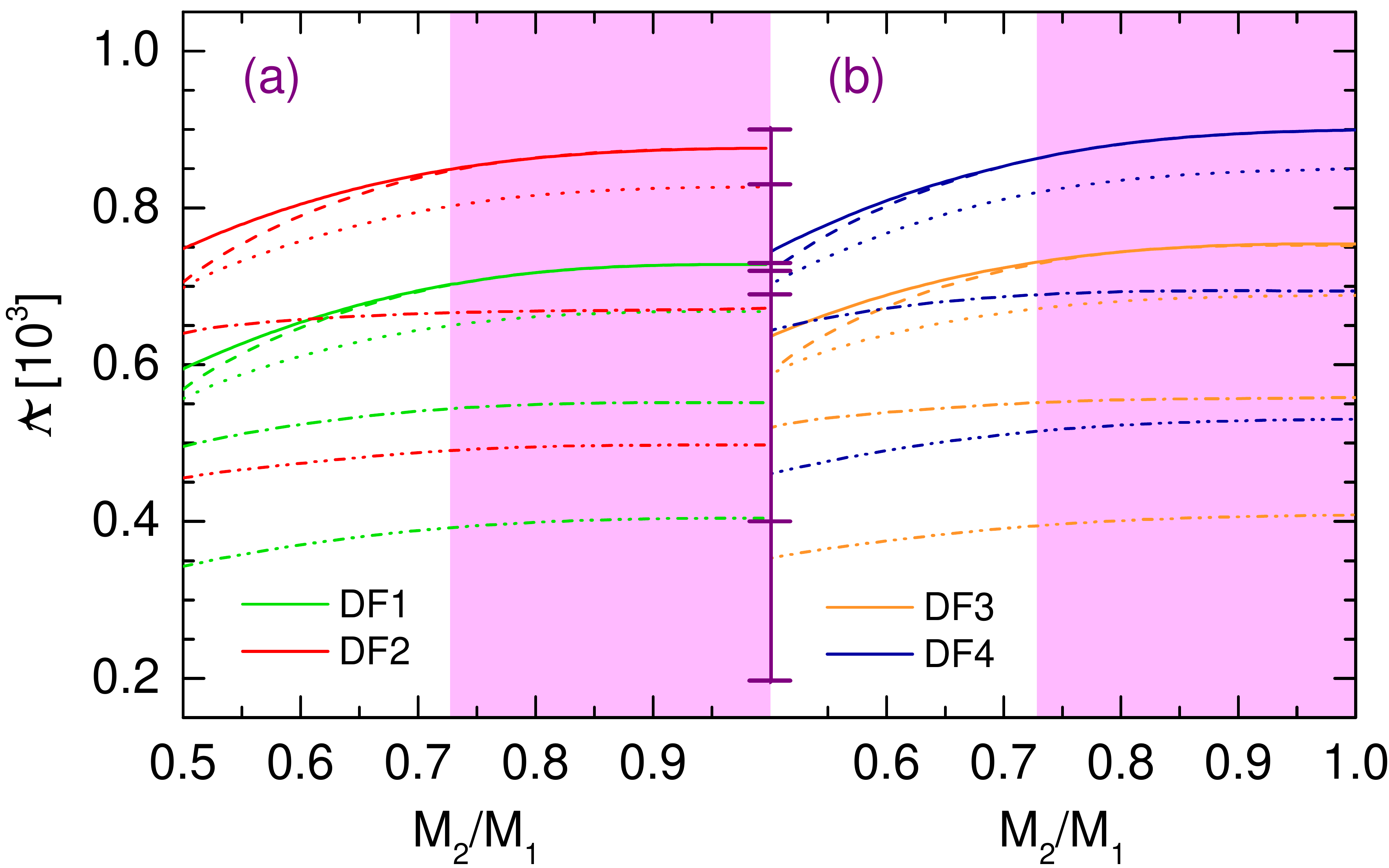}
\else
\includegraphics[width = 0.45\textwidth]{WLam_Mass.eps}
\fi
\caption{
Mass-weighted TDs of the binary as a function of the mass 
ratio, assuming a chirp mass of $\mathcal{M} = 1.186 M_\odot$. 
The error bars indicate the constraints estimated from the
GW170817 event~\citep{LIGO_Virgo2017c,LIGO_Virgo2018b}
and the electromagnetic transient AT2017gfo~\citep{Radice2018,Coughlin2018}.
The shadings show the mass ratio of the binary with 90\% 
confidence, $q \gtrsim 0.73$~\citep{Coughlin2018,LIGO_Virgo2018b}.}
\label{fig:Wlam_Mass}
\end{figure}

For the sake of completeness, in Fig.~\ref{fig:Wlam_Mass} we show the
mass weighted average TD $\tilde{\Lambda}$ with the ratio of masses of
merger components for a fixed chirp mass
$\mathcal{M} = 1.186 M_\odot$. It is seen that $\tilde{\Lambda}$
depends weakly on the mass ratio $q$. The boundaries on
$\tilde{\Lambda}$, which were set by the GW and electromagnetic
spectrum observations, respectively, are also shown. We should mention
that the estimates of both boundaries are strongly model
dependent~\citep{LIGO_Virgo2017a,
  LIGO_Virgo2018b,Radice2018,Coughlin2018,Most2018}, especially
the lower limit. The lower limit calculated from the $\Delta$-isobar
featuring EoS models is about 350. Further reduction of the radius by
up to 2~km can be obtained for larger values of  $V_\Delta$,
see~\citep{Lijj2018b} for a detailed discussion. In turn, 
$\tilde{\Lambda}$ could be decreased in this case.

%---------------------------------------------------------------
\section{Summary}
%---------------------------------------------------------------

In this work, we have explored the TDs of CSs featuring
hypernuclear matter with an admixture of $\Delta$ isobars. 
As a consequence of including $\Delta$ isobar degrees of
freedom, the dimensionless TD $\Lambda_{1.4}$ for canonical 
mass stars is reduced by about 300 for reasonably attractive 
$\Delta$-potential. Thus, {\it the presence of $\Delta$ isobars
lifts the tension between the predictions of the hypernuclear 
density functionals (which predict large TD) and 
the observations, which imply small TD.}

In addition, we found that the nuclear characteristics $Q_{\text{sat}}$
and $L_{\text{sym}}$ control the TDs of large- and small-mass stars,
respectively, whereas they  are equally important for the TDs of canonical
mass $M \simeq 1.4M_\odot$ stars. Thus, one may conclude that the
GW170817 event is highly useful in constraining these universal
parameters.

\acknowledgments{ J.L. is supported by the Alexander von Humboldt
  Foundation.  A.S. acknowledges the support by the DFG (grant No. SE
  1836/4-1), the European COST Action ``PHAROS" (CA16214) and the
  State of Hesse LOEWE-Program in HIC for FAIR.}

\end{document}